\documentstyle[12pt]{article}
\begin{document}
\begin{center}
\Large{QFT, STRING TEMPERATURE AND THE STRING PHASE  \\ 
OF DE SITTER SPACE TIME}\\ 
\vspace{3cm}
M. RAMON MEDRANO \footnotemark[1] , \footnotemark[2]
\footnotetext[1]{Departamento de F\'{\i}sica Te\'orica, Facultad de Ciencias
F\'{\i}sicas, Universidad Complutense, Ciudad Universtaria, E-28040, Madrid, Spain.} 
and N. SANCHEZ \footnotemark[2] 
\footnotetext[2]{Observatoire de Paris, Demirm (Laboratoire Associ\'e au CNRS UA 
336, Observatoire de Paris et Ecole Normale Sup\'erieure), 
61 Avenue de l'Observatoire, 75014 Paris,
France.}
\end{center}
\normalsize

\bigskip\bigskip\bigskip

\begin{abstract}
\noindent
The density of mass levels $\rho(m)$ and the critical temperature for strings in de 
Sitter space-time are found. QFT and string theory in de Sitter space are compared. A ``Dual''- transform 
is introduced which relates classical to quantum string lengths, and more generally, QFT and string 
domains. Interestingly, the string temperature in De Sitter space turns out to be the -Dual transform
of the QFT-Hawking-Gibbons temperature.

\medskip
\noindent
The back reaction problem for strings in de Sitter space is addressed selfconsistently in the framework
of the ``string analogue'' model (or thermodynamical approach), which is well suited to combine QFT and 
string study. We find de Sitter space-time is a self-consistent solution of the semiclassical Einstein
equations in this framework. Two branches for the scalar curvature $R(\pm)$ show up: a 
classical, low curvature solution $(-)$, and a quantum high curvature solution $(+)$, enterely
sustained by the strings. There is a maximal value for the curvature $R_{\max}$ due to the
string back reaction.

\medskip
\noindent
Interestingly, our Dual relation manifests itself in the back reaction solutions: the $(-)$ branch is
a classical phase for the geometry with intrinsic temperature given by the QFT-Hawking-Gibbons 
temperature. The $(+)$ is a stringy phase for the geometry with temperature given by the intrinsic
string de Sitter temperature. $2 + 1$ dimensions are considered, but conclusions hold generically
in $D$ dimensions.
\end{abstract}

\newpage

\section{INTRODUCTION AND RESULTS}

\bigskip\bigskip
In the context of Quantum Field Theory in curved spacetime, de Sitter spacetime has a Hawking-Gibbons 
temperature given by

\bigskip

$$T_{DS} = \frac{\hbar}{2\pi K_B} H = \frac{\hbar c}{2 \pi K_B} \: \frac{1}{L_{DS}} $$

(see Ref. [1] for its appropriated interpretation), $H$ being the Hubble constant, 
$L_{DS} = c H^{-1}$ being the classical horizon size.

\bigskip

In the context of String Theory in curved spacetime, strings in de Sitter spacetime have a maximal 
or critical temperature given by

\bigskip

$$T_S = \frac{c^3}{\alpha^\prime K_B H} = \frac{\hbar c}{K_B} \: \left( \frac{L_{DS}}{L_S^2} \right)$$

(see section $3$ in this paper for its appropriated derivation), 
$L_S \equiv \left( \frac{\alpha^\prime \hbar}{c} \right)^{\frac{1}{2}}$ being a characteristic string
length scale.

We introduce here a ${\cal R}$- or ``Dual'' -transformation over a length:

\bigskip

$$\tilde{L} = {\cal R} L = L_{\cal R}^2 L^{-1} $$

ie, if $L \equiv L_{DS}$, then:

\bigskip

$$\tilde{L}_{DS} = \frac{\alpha^\prime \hbar}{c^2} \: H  $$

\bigskip
$\tilde{L}_{DS}$ is precisely the Compton length of a particle whose mass is given by

\bigskip

$$m_{\max} = \left( \frac{c}{\alpha^\prime H} \right)   $$

\bigskip
This is the maximal mass for the spectrum of particle (oscillating or stable) string states in de Sitter
spacetime Ref. [2], [3] and [4].

The ${\cal R}$ -transfromation links classical lengths to quantum string lengths. (In de Sitter 
spacetime, it links the classical horizon size $L_{DS}$ to the quantum string size in this spacetime. 
We are refering here to the oscillatory or stable strings (those from which the quantum particle
states derive)).

\bigskip
Under the ${\cal R}$ -transformation (see section $3$):

\bigskip

$$T_S = 2 \pi \tilde{T}_{DS}    $$

\bigskip
The string temperature in de Sitter spacetime turns out $2 \pi$ times the ``Dual'' (${\cal R}$ 
-transformated) of the Hawking temperature (and conversely). That is, the intrinsic QFT and string
temperatures in de Sitter space are ${\cal R}$ -Dual one of each other. In fact, this has a more
general validity: the ${\cal R}$~-transform can maps QFT and string domains (or regimes) and applies 
to other spacetimes as well. In particular, it plays a key role when applied to black holes Ref. [5].

\bigskip
In the context of QFT, de Sitter (as well as AdS) spacetime, is an exact solution of the semiclassical
Einstein Equations with back reaction included Ref. [6] and [7]. Semiclassical in this context
means that quantum matter fields (including the graviton) are coupled to $c$-number gravity and the 
vacuum expectation value of matter energy momentum tensor acts in turn as a source of gravity (quantum
back reaction effect). \\

In this paper we investigate the quantum back reaction effect of strings in de Sitter spacetime. In 
principle, this question should be properly addressed in the context of String Field Theory. On the 
lack of a tractable framework for it, we work here in the framework of the string analogue model (or 
thermodynamical approach): the string as a collection of fields $\Phi_n$ coupled to the classical 
background, and whose masses $m_n$ are given by the degenerate string mass spectrum in the curved
space considered. (The fields $\Phi_n$ are without self-interaction but are coupled to the classical 
geometry). The fields $\Phi_n$ are ``repeated'' $\rho (m)$ times, the degeneracy of states being given 
by $\rho (m)$, the density of mass levels of the string.

\bigskip
In flat spacetime, the higher masses string spectrum is given by

\bigskip

$$\rho (\bar{m}) \simeq \bar{m}^{-a} e^{b \bar{m}} \: , \: \bar{m} \equiv 
\sqrt{\frac{\alpha^\prime c}{\hbar}} \: m $$

($a$ and $b$ being constants, depending on the model, and on the number of space dimensions).

\bigskip
In de Sitter spacetime, we find $\rho (m)$ is given by [Eq. (33.b)]:

\bigskip

$$\rho (\bar{m}) = \frac{\bar{m}}{\Gamma} \: \frac{4 \gamma^2}{(1 - \Gamma)^2} \exp \lbrack 
\frac{4 \pi^2}{3 \gamma} \: (1 - \Gamma) \rbrack^\frac{1}{2} $$

\bigskip

$$\Gamma \equiv (1 - \bar{m}^2 \gamma)^\frac{1}{2} \: , \: \gamma \equiv \frac{5 \alpha^\prime \hbar}
{4 c^3} H^2$$

\bigskip
It satisfies the behaviour

\bigskip

$$\rho (\bar{m}) \sim e^{\frac{2 \pi}{6} \sqrt{\frac{\alpha^\prime c}{\hbar}} m \lbrack 1 - 
\frac{5}{32} \left( \frac{m \alpha^\prime H}{c} \right)^2 + 0 \left( \frac{m \alpha^\prime H}
{c} \right)^3 \rbrack }
$$

When $H = 0$, it yields the flat spacetime asymptotic behaviour.

\bigskip
Here we deal with $2 + 1$ dimensions, but the results are the same for $D$ dimensions, only the 
numerical values of the constants will change.

In QFT, the expectation value of the $(2 + 1)$ dimensional energy-momentum tensor for a quantum
massive field (in the de Sitter invariant (Bunch-Davies) vacuum) Ref. [8] and [9] is given by
[Eq. (35)]. In the framework of the analogue model, the string vacuum expectation value $<\tau_\mu^\mu>$
is given by

\bigskip

$$\langle \tau^\mu_\mu  \rangle = \frac{\int^{m_{\max}}{m_0} \langle T^\mu_\mu (m) \rangle_S \rho 
(m) }{\int^{m_{\max}}_{m_0} \rho (m) dm} $$

$<T^\mu_\mu (m)>_S$ being the trace stress tensor vacuum expectation value for an individual quantum 
field with mass in the string mass spectrum. $m_0$ is the lower mass from which the asymptotic 
expression for $\rho (m)$ is still valid.

\bigskip

We apply self consistently the string $<\tau^\mu_\mu >$ to the rhs of the semiclassiacl Eisntein 
Equations, we study the back reaction effect in de Sitter space of the higher excited string modes.
In constant curvature spaces (such as dS and AdS) the semiclassical back reaction equations yield the 
scalar curvature in terms of $H$ and of the quantum matter content (the trace $<\tau^\mu_\mu >$).

\bigskip
The mass domain for fields in de Sitter spacetime is given by

\bigskip

$$m_{\rm QFT} < \frac{\hbar H}{c^2} $$

while in string theory, the string mass in de Sitter spacetime satisfies

\bigskip

$$m_S < \frac{c}{\alpha^\prime H} $$

\bigskip

Under the ${\mathcal R}$ -transformation we have

\bigskip

$$
\begin{array}{rcccl}
\tilde{m}_{\rm QFT} &=& {\cal R} m_{\rm QFT} &=& m_S    \\  
{\tilde{m}}_S &=&  {\cal R} m_S &=& m_{\rm QFT}      \\
{\cal R} <T^\mu_\mu >_{\rm QFT} &=& <\tilde{T}^\mu_\mu >_{\rm QFT} 
&\equiv & <T^\mu_\mu >_S
\end{array}
$$

\bigskip
here $<T^\mu_\mu >_S$ is given by [Eq. (46)] as a function of the variable $x \equiv \left(
\frac{m}{m_{\max}} \right)^2$ .

\bigskip
We find $<\tau^\mu_\mu >$ up to order $\gamma$ (as given by [Eq. (55.b)]), in terms of 
$\alpha^\prime$ and of the scalar curvature $R = \frac{6 H^2}{c^2}$:

$$<\tau^\mu_\mu > = - \frac{\hbar H^3}{3 \pi^3 c^2} \sqrt{6 \gamma} \left( 1 + \frac{2}{\pi}
\sqrt{6 \gamma} \right) $$

\bigskip
Inserting it selfconsistently in the semiclassical Einstein equations for the effective
geometry, we find for the scalar curvature:

\bigskip

$$R_\pm = 6 \Lambda_\pm = \frac{1}{2} R_{\max} \lbrack 1 \pm \left( 1 - 4\frac{R}{R_{\max}}
\right)^{\frac{1}{2}} \rbrack $$

\bigskip
Due to the quantum string back reaction, the curvature reachs a maximum value:

\bigskip

$$R_{\max} = \frac{9 c^4 \pi^2}{4 G} \: \left( \frac{6}{5 \alpha^\prime c \hbar^3} 
\right)^\frac{1}{2} $$

\bigskip
Three cases show up depending on wether $(i)$ $R < \frac{1}{4} R_{\max}$, $(ii)$ $R = \frac{1}
{4} R_{\max}$ or $(iii)$ $R < \frac{1}{4} R_{\max}$.

Case $(i)$ describes two semiclassical de Sitter spacetimes with constant positive curvatures 
$R_{\pm} > 0$ and well defined associated temperatures $T_{(\pm )}$ .

Case $(ii)$ describes one semiclassical de Sitter (positive curvature) space for which 
$R_+ = R_- = \frac{1}{2} R_{\max}$.

For $(iii)$, no real spacetime geometries, nor temperatures are possible.

\bigskip
Two branches, $(+)$ and $(-)$, for the curvature show up. The leading term is $R$ in the $(-)$ 
branch, while is $R_{\max}$ in the $(+)$ branch. In an expansion in $\frac{R}{R_{\max}}$, 
classical de Sitter space is recovered in the $(-)$ branch. $R (-)$ is a low curvature, 
classically allowed solution, while $R (+)$ is a ``quantum'' branch (it does not exists 
classically) and its curvature is very high. The quantum string back reaction generates this
branch.

Our Dual relation between classical-QFT and the string domains manifests here again in the back
reaction solutions: the branch $(-)$ is a classical phase for the geometry which temperature is 
given by the QFT Hawking-Gibbons temperature $T (-) = T_{DS} = \frac{\hbar c}{2 \pi K_B} 
\left( \frac{R_-}{6} \right)^{\frac{1}{2}}$.

The branch $(+)$ is a stringy phase for the geometry which temperature is the intrinsic string
de Sitter temperature $T (+) = T_S = \frac{c^2}{\alpha^\prime K_B} \left( \frac{6}{R_+} 
\right)^{\frac{1}{2}}$.

\bigskip
Moreover, our Dual relation and the two phases: a classical-QFT phase (with the Hawking 
temperature) and a quantum-string phase (with the string temperature), appear to be a generic
feature and are very enlighting for black holes. Our study of the string black hole temperature
and quantum string back reaction for black holes is reported elsewhere Ref. [5].

\bigskip
This paper is organized as folows: In Section 2 we summarize de Sitter spacetime and 
the QFT Hawking-Gibbons temperature. In Section 3, we derive the string temperature
in de Sitter spacetime and its Dual-relation to the Hawking-Gibbons temperature. In section
4 we find the quantum string back reaction and its solution. In Section 5
we present the concluding remarks.

\newpage

\section{De Sitter space-time} 

\bigskip
\bigskip
\hspace{0.5cm} De Sitter space-time is a cosmological space with constant scalar curvature ($R$), and 
vanishing spatial curvature index ($K$). Its D-dimensional metric is given by \\

\begin{eqnarray}
ds^2 & = & - c^2 dt^2 + a^2 (t) (dr^2 + r^2 d \Omega^2_{D-2})  \\    
a(t) & = & e^{Ht}     \nonumber
\end{eqnarray}

\bigskip
\noindent
$t$ being the cosmic time and $H = \frac{d \ln a(t)}{dt}$ being the Hubble constant. \\

De Sitter space-time can be generated by a cosmological constant ($\Lambda$). The curvature 
$R_{\mu\nu}$, $R$, $H$ and $\Lambda$ are related by

$$ R_{\mu\nu} - \frac{1}{2} g_{\mu\nu} R + \Lambda g_{\mu\nu} = 0  \eqno(2.a)$$

\bigskip
$$ R = D(D-1) \frac{H^2}{c^2} = \frac{2D}{D-2} \: \Lambda  \eqno(2.b)$$

\bigskip
$$\Lambda = \frac{(D-1)(D-2)}{2} \: \frac{H^2}{c^2}   \eqno(2.c)$$

\bigskip
The D-dimensional de Sitter metric can be also expressed in terms of the so-called static
coordinates

\setcounter{equation}{2}
\begin{equation}
ds^2 = - A(r) c^2 dT^2 + A^{-1} (r) dr^2 + r^2 d\Omega^2_{D-2}  
\end{equation}

\noindent
where \\

\begin{equation}
A(r) = 1 - \frac{H^2 r^2}{c^2}   
\end{equation}

\bigskip
\noindent 
which show the existence of a horizon at \\

\begin{equation}
r = L_{DS} = c H^{-1}   
\end{equation}

\bigskip
In the context of Quantum Field Theory in curved space-time, de Sitter space-time has a 
Hawking-Gibbons temperature ($T_{DS}$) given by Ref. [1]

$$ T_{DS} = \frac{H \hbar}{2 \pi K_B}   \eqno(6.a) $$

\bigskip
\noindent
Notice that this expression for $T_{DS}$ holds in any number of space-time dimensions. \\
In terms of $R$ and $\Lambda$, $T_{DS}$ reads as

$$T_{DS} = \frac{\hbar c}{2\pi K_B} \: \sqrt{\frac{R}{D(D-1)}}   \eqno(6.b) $$

\noindent
or
\bigskip

$$ T_{DS} = \frac{\hbar c}{2\pi K_B} \: \sqrt{\frac{2\Lambda}{(D-1)(D-2)}}   \eqno(6.c) $$

\bigskip
\bigskip
If one defines a surface gravity $\kappa_{DS}$ equal to $cH$, $T_{DS}$ [Eq. (6)] reads

\setcounter{equation}{6}
\begin{equation}
T_{DS} = \frac{\hbar \kappa_{DS}}{2\pi K_B c}   
\end{equation}

\bigskip
Furthermore, $T_{DS}$ can be also expressed in terms of the classical length scale $L_{DS}$ [Eq. (5)]
as

\begin{equation}
T_{DS} = \frac{\hbar c}{2\pi K_B} \: \frac{1}{L_{DS}}   
\end{equation}

\bigskip \bigskip \bigskip

\section{Quantum strings in de Sitter space-time}  

\bigskip
\bigskip
\hspace{0.5cm} String theory in de Sitter space-time is exactly integrable in any dimension Ref. [10]. 
However, explicit expressions for the string solutions are not easy to write due to the complexity 
of the equations, and even of the solutions. Two main quantum frameworks have been studied and solved~: 
$(i)$ canonical quantization of generic strings in any dimension Ref. [2] and [3]; $(ii)$ 
semiclassical quantization of exact circular strings configurations in $2+1$ space-times Ref. [4].  \\

In this section we will consider the case $(i)$ (canonical quantization). We remind now the reader 
of some of the main issues [Eq. (2)] and [Eq. (3)].  \\
\hspace{0.5cm} In canonical quantization, one treats the de Sitter classical background exactly, and 
considers the
string oscillations around its center of mass as perturbations. The string center of mass is an exact 
solution of the geodesic equation. The perturbations (dimensionless) parameter is here

\bigskip
\begin{equation}
\frac{L_S}{L_{DS}} = \frac{H}{c} \: \sqrt{\frac{\alpha^{\prime} \hbar}{c}} = \sqrt{\frac{2 \Lambda 
\alpha^{\prime} \hbar}{(D-1)(D-2)c}} \ll 1    
\end{equation}

\bigskip
\noindent
where $L_{DS}$ (de Sitter length or horizon) is given by [Eq. (5)] and $L_S$ (string length scale) is

\begin{equation}
L_S \equiv \left( \frac{\alpha^{\prime} \hbar}{c}\right)^{\frac{1}{2}}   
\end{equation}

\bigskip
\noindent
Here $\alpha^{\prime} \equiv \frac{c^2}{2 \pi T}$, where $T$ is the string tension 
($(\alpha^{\prime})^{-1}$ : linear mass density).  \\

In this framework, the mass spectrum formula in de Sitter space for an $N -{\rm th}$ level state is 
given by Ref. [2] and [3]

\begin{equation}
\alpha^{\prime} \left( \frac{c}{\hbar}\right) m^2 = 24 \sum_{n > 0} \frac{2n^2 - H^2 m^2   
\frac{\alpha^{\prime 2}}{c^2}}{\sqrt{n^2 - H^2 m^2  \frac{\alpha^{\prime 2}}{c^2}}} +2N           
\frac{2 - H^2 m^2  \frac{\alpha^{\prime 2}}{c^2}}{\sqrt{1 - H^2 m^2  \frac{\alpha^{\prime 2}}{c^2}}}
\end{equation}

\bigskip
One of the consequences of the spectrum is that the number of string oscillating states, although
being very large, is finite. This maximum number is given by  \\

\begin{equation}
N_{\rm max} \simeq {\rm Int} \left\lbrack 0 \cdot 15 \left( \frac{L_{DS}}{L_S} \right)^2\right\rbrack 
= {\rm Int} \left\lbrack 0 \cdot 15 \left( \frac{c^3}{\alpha^{\prime} \hbar H^2} \right) \right\rbrack              
\end{equation}       

\bigskip
\noindent
Furthermore, there is a maximum mass ($m_{\rm max}$) for the corresponding real mass solutions   \\

\begin{equation}
\alpha^{\prime} \left( \frac{c}{\hbar}\right) m^2_{\rm max} \simeq \left( \frac{L_{DS}}{L_S}\right)^2        
\end{equation}

\noindent
ie $$m^2_{\rm max} \simeq \left( \frac{c}{\alpha^{\prime} H}\right)^2$$

\bigskip
\noindent
As [Eq. (9)] is fulfilled for oscillating string states, ie $1 \ll \left( \frac{L_S}{L_{DS}} \right)^2
\left( = \frac{c^3}{\alpha^{\prime} H^2 \hbar}\right)$~, the number of oscillating strings and the 
maximum string mass are large.  \\

The fact that there is a maximum mass implies the existence of a maximum or critical temperature for 
the strings in de Sitter space-time. The temperature $T_S$ corresponding to $m_{\max}$ [Eq. (13)], is 
given by  \\

\begin{equation}
T_S = \frac{c^3}{\alpha^{\prime} H K_B}         
\end{equation}

\bigskip
\noindent
Or, in terms of the classical and string length scales $L_{DS}$ and $L_S$~:

\bigskip
\begin{equation}
T_S = \frac{\hbar c}{K_B} \left( \frac{L_{DS}}{L^2_S}\right)      
\end{equation}

\bigskip
If we compare this maximal or critical temperature for strings in de Sitter space-time ($T_S$)
with the quantum field theory Hawking-Gibbons temperature for de Sitter space-time ($T_{DS}$) 
[Eq. (6.a)], we have  \\

\begin{equation}
T_S = \left( \frac{c^3 \hbar}{2 \pi \alpha^{\prime} K^2_B}\right) \frac{1}{T_{DS}} \: \cdot    
\end{equation}

\bigskip
\bigskip
\noindent
Let us define now the following transformation $\cal R$ over a length $L$  \\

\begin{equation}
\tilde{L} = {\cal R} L = L^2_{\cal R} L^{-1}      
\end{equation}

\bigskip
If $L_{\cal R} = L_S$ [Eq. (10)], and we apply this transformation to $L \equiv L_{DS}$
[Eq. (5)], we obtain  \\

\begin{equation}
\tilde{L}_{DS} = {\cal R} L_{DS} = \frac{\alpha^{\prime} \hbar H}{c^2}   
\end{equation}

\bigskip
\noindent
But $\tilde{L}_{DS}$ is precisely the (reduced) Compton wave length ($\lambda = 
\frac{\hbar}{mc}$) of a particle whose mass is equal to $m_{\rm max}$ given by [Eq. (13)].
ie $\tilde{L}_{DS}$ is the minimal quantum length of a string in de Sitter space. Therefore, this
transformation links the classical de Sitter length scale $L_{DS}$ to the quantum string- de Sitter
one, $\tilde{L}_{DS}$. \\

\hspace{0.5cm} The string temperature $T_S$ [Eq. (15)] in de Sitter space time can be rewritten in 
terms of $\tilde{L}_{DS}$ [Eq. (18)] as  \\

\begin{equation}
T_S = \frac{\hbar c}{K_B} \: \frac{1}{\tilde{L}_{DS}}       
\end{equation}

\bigskip
We see now from [Eq. (8)] and [Eq. (19)] that the following relations hold under the ${\cal R}$
transformation \\

\begin{equation}
\tilde{T}_{DS} = \frac{1}{2 \pi} \: T_S      
\end{equation}

\noindent
and

\begin{equation}
\tilde{T}_S = 2 \pi \: T_{DS}      
\end{equation}

\bigskip
\noindent
From the above equations we can read as well  \\

$$T_S \: T_{DS} = \tilde{T}_S \: \tilde{T}_{DS}$$

\bigskip
That is, the maximal string temperature in de Sitter spacetime is the Dual (in the sense of the 
$\cal{R}$- transformation [Eq. (17)]) of the Hawking (QFT) temperature.

\bigskip \bigskip \bigskip

\section{Quantum String back reaction in de Sitter space-time}

\bigskip \bigskip
When quantum matter (particle fields, strings) is present in de Sitter space-time, the relation 
between the scalar curvature and the cosmological constant $\Lambda$ will be modified through the 
semiclassical Einstein equations. Semiclassical in this context means that matter, which is a q-number,
is coupled to c-number gravity through the equations  \\

\begin{equation}
R_{\mu\nu} - \frac{1}{2} \: R \: g_{\mu\nu} + \Lambda g_{\mu\nu} = \frac{8 \pi G}{c^4} 
\left\langle \tau_{\mu\nu} (q,g_{\mu\nu}) \right\rangle        
\end{equation}

\bigskip
\noindent
The space-time background metric $g_{\mu\nu}$ generates a non-zero vacuum expectation value of the 
energy momentum tensor $\left\langle \tau_{\mu\nu} \right\rangle$, which in turn acts as a source of 
curvature (For instance, in 4-dimensional Quantum Field Theory, matter fields $\hat{\phi}$    
include the graviton and $\left\langle T_{\mu\nu} \left( \hat{\phi} , g_{\mu\nu}\right) \right\rangle$
is calculated up to one loop order, where $\langle T_{\mu\nu} \rangle$ stands for its renormalized 
value Ref. [11] and [7]).  \\

For maximally symmetric (constant curvature) spaces (such as de Sitter and anti de Sitter), these
equations read  \\

\begin{equation}
\left( \frac{2-D}{2D} \: R + \Lambda \right) g_{\mu\nu} = \frac{8 \pi G}{c^4} \: \left\langle 
\tau_{\mu\nu} \right\rangle           
\end{equation}

\bigskip
\noindent
which yields the trace equation

$$R - \frac{2D}{D-2} \: \Lambda = - \frac{16 \pi G}{c^4 (D-2)} \: \left\langle \tau^\mu_\mu
\right\rangle       \eqno(24.a)$$

\noindent
Or
\bigskip
$$R = \frac{2D}{D-2} \: \Lambda_{\rm eff}    \eqno(24.b)$$

\noindent
where 
\bigskip
$$\Lambda_{\rm eff} = \Lambda - \frac{8 \pi G}{D c^4} \: \left\langle \tau^\mu_\mu \right\rangle
 \eqno(24.c)$$

\bigskip
\noindent
which shows clearly quantum matter as a source of curvature and of temperature [Eq. (7.b)]. \\

As $\left\langle \tau_{\mu\nu} \right\rangle$ is proportional to $g_{\mu\nu}$, de Sitter ($DS$) 
(as well as anti de Sitter ($A_{DS}$) backgrounds are exact self consistent solutions to [Eq. (22)]
 with back reaction included.  \\

In de Sitter space, there is one real parameter $\alpha$-- family of de Sitter group invariant vacua
$ \vert \alpha >$. Here $\left\langle \tau_{\mu\nu} \right\rangle$ is the expectation value in the
Bunch-Davies Ref. [8] and [9] (``euclidean'' or ``inflationary'') vacuum obtained for $\alpha 
= 0$.  \\

\bigskip 

In order to study the back reaction problem for string theory in de Sitter space-time, we will work 
in the framework of the string analogue model, and in a $2 + 1$ space-time, where we will use the
results coming from semiclassical quantization of exact circular strings configurations Ref. [4].
However, one should have in mind that the $2 + 1$ string dynamics could be embedded in a higher 
dimensional space-time and our results generalized to higher dimensions as well. \\

In the spirit of the analogue model, we consider here the string as a collection of fields
$\hat\phi_n$ coupled to the classical background, and whose masses $m_n$ are given by
the degenerate mass spectrum of the string. The fields $\hat{\phi}_n$ are free (without self 
interactions) but interact here with the classical geometry. The (higher) mass spectrum is described by
the density of mass levels $\rho (m)$. As it is known, in flat space-time $\rho (m)$ is given by \\

\setcounter{equation}{24}

\begin{equation}
\rho (\bar{m}) \sim \bar{m}^{-a} \exp b \: \bar{m}               
\end{equation}

\bigskip
\noindent
where we have introduced the adimensional mass variable

\begin{equation}
\bar{m} \equiv \sqrt{\frac{\alpha^\prime c}{\hbar}} \: m        
\end{equation}

\bigskip
\noindent
(which will prove useful later on.) The constants $a$ and $b$ depend on the string model and on the
dimensions of the space-time.  \\

In de Sitter space-time, $\rho (m)$ has a different behavior from the one of [Eq. (25)], as it follows
from the string mass spectrum in de Sitter space [Eq. (11)]. \\

Classical string equations of motion and constraints have been solved exactly for circular string 
configurations $(t = t(\tau) , r = r (\tau) , \phi = \sigma )$ in a $2 + 1$ de Sitter 
space-time Ref. [12]. Semiclassical quantization of the time periodic (oscillating) solutions has been 
performed Ref. [4]. For $\frac{\alpha^\prime H^2 \hbar} {c^3} \ll 1 \: ( {\rm ie} \left( \frac{L_S}
{L_{DS}} \right)^2 \ll 1 )$ , corresponding to the semiclassical quantization here, 
and which is always satisfied for oscillating strings, the results are the following~: $(i)$ 
the quantized mass formula is given, for large $n$ , by  \\

\begin{equation}
\alpha^\prime \left( \frac{c}{\hbar } \right) m^2 \simeq 4n (1 - \gamma n)       
\end{equation}

\noindent
where

\begin{equation}
\gamma \equiv \frac{5 \alpha^\prime H^2 \hbar}{4c^3}             
\end{equation}

\bigskip
\noindent
(Notice that for $H = 0$ one recovers the mass formula for closed strings in Minkowski space); $(ii)$ 
the number of oscillatory circular string states, although being very large, is finite  \\

\begin{equation}
N_{\rm max} \simeq {\rm Int} \left\lbrack 0 \cdot 34 \frac{c^3}{\alpha^\prime H^2 \hbar} \right\rbrack
\end{equation}

\bigskip
\noindent
and $(iii)$ the level spacing is approximately constant, in $\left( \frac{\alpha^\prime c}{\hbar} 
\right)^{-1}$ units (although smaller than in Minkowski space-time and slightly decreasing).  \\

\bigskip
Furthermore, from [Eq. (27)] and [Eq. (28)], the maximum value for the string mass states is given by \\

$$ m^2_{\rm max} \simeq \frac{4}{5} \left( \frac{c}{\alpha^\prime H} \right)^2      \eqno(30.a) $$

\noindent
or
\bigskip

$$ \bar{m}^2_{\rm max} \simeq \gamma - 1       \eqno(30.b) $$

\bigskip
\noindent
(See [Eq. (26)] and [Eq. (28)]). The above results are in very good agreement with the ones 
corresponding to canonical quantization of generic strings [Eq. (12)] and [Eq. (13)].  \\

It must be noticed that [Eq. (30)] will provide a maximum string temperature similar to the one of 
[Eq. (14)]. \\

The asymptotic degeneracy of levels $d_n$ (in flat as well as in curved space-time) is generically
$\sim n^{- \frac{(D + 1)}{2}} e^{4 \pi \sqrt{\frac{(D - 2)n}{6}}}$ for any non-compact $D-$ dimensional
space-time. For closed string solutions and $D = 3$ , the asymptotic degeneracy of levels $d_n$ reads

$$d_n \sim n^{-2} e^{4\pi\sqrt{\frac{n}{6}}}     \eqno(31.a)  $$

\bigskip
\noindent
where $n$ has now to be expressed as a function of the quantized mass. It is through the relation 
$m = m(n)$ of the mass spectrum, that the differences due to the space-time curvature enter in the 
above formula.  \\

The density of mass levels $\rho (m)$ and the degeneracy $d_n$ satisfy

$$\rho (m) d m = d_n (m) d n       \eqno(31.b) $$


From [Eq. (27)] and [Eq. (28)] we have

\bigskip
$$n \simeq {\rm Int} \left\lbrace \frac{2 c^3}{5 \alpha^\prime H^2 \hbar} \left\lbrack 1 - \left( 1 -
\frac{5}{4} \left( \frac{\alpha^\prime H m}{c} \right)^2 \right)^{\frac{1}{2}} \right\rbrack 
\right\rbrace           \eqno(32.a)  $$

\noindent
Or 
\bigskip

$$n \simeq {\rm Int} \left\lbrace \frac{1}{2\gamma} \left\lbrack 1 - \left( 1 - \bar{m}^2 \gamma
\right)^{\frac{1}{2}} \right\rbrack \right\rbrace \: \: \: ,          \eqno(32.b)  $$

\bigskip
\noindent
in terms of the adimensional variables $\bar{m}$ and $\gamma$ [Eq. (26)] and [Eq. (28)].  \\

Therefore from [Eq. (31.b)] and [Eq. (32.b)], the asymptotic string density of mass levels in de Sitter
space is \\

$$\rho (m) \sim \left( \alpha^\prime \frac{c}{\hbar} \right) m \: \frac{d_n}
{1 - 2\gamma n} \eqno(33.a)  $$

\bigskip
\noindent
which for $H = 0$ ($\gamma = 0$ , [Eq. (28)]) gives the flat space-time relation $\rho (m) \sim m
d_n (m)$.  \\

\medskip
From [Eq. (32.b)], [Eq. (33.a)] and [Eq. (31.a)], we obtain   \\

$$\rho (\bar{m} ) \sim \bar{m} (1 - \bar{m}^2 \gamma )^{-\frac{1}{2}} \cdot \left\lbrack 
\frac{1}{2 \gamma} \left( 1 - (1 - \bar{m}^2 \gamma)^{\frac{1}{2}} \right) \right\rbrack^{-2}
\cdot \exp \left\lbrace \frac{4\pi}{\sqrt{6}} \left\lbrack \frac{1}{2\gamma} \left( 1 - (1 - \bar{m}^2
\gamma )^{\frac{1}{2}} \right) \right\rbrack^{\frac{1}{2}} \right\rbrace       \eqno(33.b)   $$

\bigskip
\noindent
where $\gamma$ is given by [Eq. (28)]. \\

\medskip
[Eq. (33.b)] generalizes to de Sitter spacetime the standard flat space time behaviour [Eq. (25)].

\bigskip 
If we develop the exponent of $\rho (\bar{m})$ in powers of $\bar{m}^2 \gamma = \left( 
\frac{m}{m_{\rm max}} \right)^2 < 1$ [Eq. (26)] and [Eq. (30.b)], we have \\

$$\rho (\bar{m}) \: \alpha \: e^{\frac{2\pi}{6} \sqrt{\frac{\alpha^\prime c}{\hbar}} \: m \:
\left\lbrack 1 - \frac{1}{8} \left( \frac{m}{m_{\rm max}} \right)^2 + \: O \: \left\lbrack \left( 
\frac{m} {m_{\rm max}} \right)^3 \right\rbrack \right\rbrack} $$

\bigskip
\noindent
Or, showing the explicit dependence on $H$
\bigskip

$$\rho (\bar{m}) \: \alpha \: e^{\frac{2\pi}{6} \: \sqrt{\frac{\alpha^\prime c}{\hbar}} \: m \: 
\left\lbrack 1 - \frac{5}{32} \left( \frac{\alpha^\prime H}{c} \right)^2 m^2 + \: \cdots \:
\right\rbrack}  $$

\bigskip
\bigskip
\noindent
We see that for $H = 0$ one recovers the flat space time asymptotic behavior.  \\

\bigskip
Now, returning to the semiclassical Einstein Equations [Eq. (24.a)], [Eq. (24.b)] and [Eq. (24.c)], 
$<\tau^\mu_\mu >$ will be the vacuum expectation value of the trace of the stress tensor for the 
collection of fields (interacting  with the background) which correspond to the string 
tower of mass states in de Sitter space.  \\

In the framework of the analogue model, the string vacuum expectation value $<\tau^\mu_\mu >$ is 
given by  \\

\setcounter{equation}{33}
\begin{equation}
<\tau^\mu_\mu > \simeq \frac{\int^{m_{\rm max}}_{m_0} \: \langle T^\mu_\mu (m) \rangle_S \: \rho 
(m) dm} {\int^{m_{\rm max}}_{m_0} \rho (m) dm}            
\end{equation}

\bigskip
\noindent
where $\langle T^\mu_\mu \rangle_S$ is the vacuum expectation value of the trace of the
stress tensor for an individual quantum field. We integrate over string field masses and divide by
the total mass degeneracy.
In fact we should have $\langle n(m) \rangle \rho (m)$ where $\langle n(m) \rangle
\sim \int^\infty_0 k^{D-2} dk$ , but this divergent contribution cancels out (as it appears as a
multiplicative factor for both numerator and denominator).

\bigskip
Since in de Sitter space-time the number of particle oscillating states is finite, the sum goes up
to $m_{\rm max}$ [Eq. (30)], (instead of up to infinity as for Minkowski space-time); $m_0$ is the lower
mass from which the asymptotic expression of the density of mass levels [Eq. (33)] is valid. Therefore,
we are studying the back reaction effect in de Sitter space-time due to the higher excited string 
modes.  \\

\bigskip
Here $\rho (m)$ [Eq. (33)] depends only on the mass as usual, therefore $\langle T^\mu_\mu \rangle_S$ 
will be chosen for our study to be the expectation value of the stress tensor for a massive scalar 
field (in the de Sitter invariant or Bunch-Davies vacuum). \\

\bigskip
The Quantum Field theory value $\langle T^\mu_\mu (m) \rangle_{\rm QFT}$ , corresponding to a scalar 
massive field in a $2 + 1$ de Sitter space-time (in the de Sitter invariant vacuum), is given by 
Ref. [13] and [9] \\

\begin{equation}
\langle T^\mu_\mu \rangle_{\rm QFT} = \frac{\hbar H^3}{4 \pi c^2} \left( \frac{m c^2}{\hbar H} \right)^2
\left\lbrack (1 - 6 \zeta) - \left( \frac{m c^2}{\hbar H} \right)^2 \right\rbrack^{\frac{1}{2}} \cdot
{\rm ctg} \: \pi \left\lbrack (1 - 6 \zeta) - \left( \frac{m c^2}{\hbar H} \right)^2 
\right\rbrack^{\frac{1}{2}}             
\end{equation}

\bigskip
\noindent
where $\zeta$, a numerical factor, is the scalar coupling ($-\frac{1}{2} \zeta R \phi$ ; conformal 
coupling~: $\zeta = \frac{1}{8}$). Notice that for a massless scalar field there is no trace anomaly
in $2 + 1$ dimensions.
This happens too also for any odd dimensional de Sitter space-time Ref. [14]. In addition, for these 
spaces, 
$\langle T^\mu_\mu \rangle$ is finite and no renormalisation procedure is needed, in contrast to the
$D = 4$ case Ref. [15].  \\

\medskip
At this point, let us analyse the mass scales in the corresponding Quantum Field 
Theory (Q.F.T.) and  string theory in de Sitter space-time. From [Eq. (32)] and [Eq. (35)], one can 
read the following domains for the field mass in Q.F.T. and in string theory ( $\zeta = 0$ for 
simplicity)~:  \\

$$m_{\rm QFT} < \frac{\hbar H}{c^2}         \eqno(36.a) $$

\noindent
and 
\bigskip

$$m_S < \frac{2}{\sqrt{5}} \frac{c}{\alpha^\prime H}         \eqno(36.b)  $$

\bigskip
\bigskip
\noindent
which can be rewritten as well as (see [Eq. (26)], [Eq. (30.a)] and [Eq. (30.b)]) \\

$$\frac{m_{\rm QFT}}{M_H} < 1    \eqno(37.a)   $$

\noindent
and
\bigskip

$$\frac{m_S}{m_{\rm max}} = \bar{m}_S \sqrt{\gamma} < 1     \eqno(37.b) $$

\bigskip
\bigskip
\noindent
Here $M_H \equiv \frac{\hbar H}{c^2}$ is the mass scale of de Sitter space. [[Eq. (37.a)] just stands
that $m_{\rm QFT}$ is a test particle field in de Sitter space], and $m_{\rm max}$ is the maximum 
value for string states in the $2 + 1$ de Sitter semiclassical quantization.

\bigskip
Now we express the above inequalities  in terms of $L_{DS}$ [Eq. (5)] and
${\tilde L}_{DS}$ , being the later the minimal Compton wave length as before. Here, according to 
$m_{\rm max}$ given by [Eq. (30.a)], ${\tilde L}_{DS}$ is

\setcounter{equation}{37}
\begin{equation}
{\tilde L}_{DS} = \frac{\sqrt{5}}{2} \: \frac{\alpha^\prime H \hbar}{c^2}      
\end{equation}

\bigskip
\noindent
(semiclassical and canonical quantizations of the string just differ in the factor $\frac{\sqrt{5}}
{2}$).

\bigskip
 From [Eq. (36.a)], [Eq. (36.b)], [Eq. (5)] and [Eq. (38)] we have

$$m_{\rm QFT} < \frac{\hbar}{c} \: \frac{1}{L_{DS}}     \eqno(39.a)   $$

\noindent
and
\bigskip

$$m_S < \frac{\hbar}{c} \: \frac{1}{{\tilde L}_{DS}}     \eqno(39.b)  $$

\bigskip 
But these domains are going to be exchanged by the ${\cal R} -$ transformation given by [Eq. (18)]. 
In fact, if we apply the ${\cal R} -$ transformation to both sides of [Eq. (38.a)] and [Eq. (38.b)] we
obtain \\

\bigskip
$${\tilde m}_{\rm QFT} < \frac{\hbar}{c} \: \frac{1}{{\tilde L}_{DS}}    \eqno(40.a) $$

\noindent
and 
\bigskip

$${\tilde m}_S < \frac{\hbar}{c} \: \frac{1}{L_{DS}}      \eqno(40.b)  $$

\bigskip
\noindent
( $L_{\cal R} = \left( \frac{\sqrt{5} \alpha^\prime \hbar}{2c} \right)^{\frac{1}{2}}$ in [Eq. (7)]. The 
numerical factor $\frac{\sqrt{5}}{2}$ , that appears here and in ${\tilde L}_{DS}$ as well, is just
due to the slightly smaller $m_{\rm max}$ one obtains in semiclassical quantization, [Eq. (30.a)], as
compared with the canonical quantization, [Eq. (13)]. Obviously, the action of the ${\cal R} -$ 
transformation is equal for both cases).  \\

\medskip
We can summarize the action of the ${\cal R} -$ transformation, on the masses and on their domains, in
the following equations [Eq. (37.a)], [Eq. (37.b)], [Eq. (40.a)] and [Eq. (40.b)]  \\

$${\tilde m}_{\rm QFT} = {\cal R} m_{\rm QFT} = m_S      \eqno(41.a) $$

\bigskip

$${\tilde m}_S = {\cal R} m_S = m_{\rm QFT}         \eqno(41.b)  $$

\bigskip

$${\cal R} \left( \frac{m_{\rm QFT} c^2}{\hbar H} \right) = \frac{m_S}{m_{\rm max}} = \bar{m}_S
\sqrt{\gamma}       \eqno(41.c)  $$

\bigskip \bigskip

Now we are able to write the v.e.v. of the stress tensor $\langle T^\mu_\mu \rangle_S$
that appears in the r.h.s. of [Eq. (34)], and which corresponds to the high masses of the string
domain. From the previous mass-domain study, it is clear that $\langle T^\mu_\mu \rangle_S$ is
precisely the ${\cal R} - $ transformed object of $\langle T^\mu_\mu \rangle_{\rm QFT}$ ie  \\

\setcounter{equation}{41}
\begin{equation}
{\cal R} \langle T^\mu_\mu \rangle_{\rm QFT} = {\tilde {\langle T^\mu_\mu \rangle}}_{\rm QFT} \equiv
\langle T^\mu_\mu \rangle_S                
\end{equation}

\bigskip \bigskip
Applying the ${\cal R} -$ transformation [Eq. (41.a)] and [Eq. (41.c)] to $\langle T^\mu_\mu 
\rangle_{\rm QFT}$ , given by [Eq. (35)], we obtain  \\

\begin{equation}
{\tilde {\langle T^\mu_\mu \rangle}} = \frac{\hbar H^3}{4 \pi c^2} ( \bar{m}^2 \gamma ) \left\lbrack
(1 - 6 \zeta ) - \bar{m}^2 \gamma \right\rbrack^{\frac{1}{2}}  \cdot {\rm ctg} \: \pi \left\lbrack
(1 - 6 \zeta ) - \bar{m}^2 \gamma \right\rbrack^{\frac{1}{2}}            
\end{equation}

\bigskip
\bigskip
\noindent
in terms of the adimensional variable  \\

$$\bar{m}^2 \gamma = \frac{5}{4} \left( \frac{\alpha^\prime m H}{c} \right)^2  $$

\bigskip
\noindent
[Eq. (26)] and [Eq. (28)].

\bigskip
\bigskip

In order to compute $\langle \tau^\mu_\mu \rangle$ [Eq. (34)], it is convenient to express $\rho 
( \bar{m} ) d \bar{m}$ [Eq. (33)] and ${\tilde {\langle T^\mu_\mu \rangle}}$ [Eq. (43)], in terms of the
adimensional variable $x \equiv m^2 \gamma$ (running ratio $\frac{m^2}{m^2_{\rm max}}$ )~:

\bigskip
\begin{equation}
\rho ( \bar{m} ) d \bar{m} = \frac{1}{2\gamma} \cdot \rho (x) dx         
\end{equation}

\bigskip
\noindent
where
\bigskip

\begin{equation}
\rho (x) \equiv (1 - x)^{- \frac{1}{2}} \cdot \left\lbrack  1 - (1 - x)^{\frac{1}{2}} 
\right\rbrack^{-2} \cdot \exp \left\lbrace \frac{2\pi}{\sqrt{3\gamma}} \left\lbrack 1 - 
(1 - x)^{\frac{1}{2}} \right\rbrack^{\frac{1}{2}} \right\rbrace               
\end{equation}

\bigskip
\noindent
and 
\bigskip

\begin{equation}
{\tilde {\langle T^\mu_\mu \rangle}} = - \frac{\hbar H^3}{4 \pi c^2} F(x)         
\end{equation}

\bigskip
\noindent
where
\bigskip

\begin{equation}
F(x) \equiv -x \: (1 - x)^{\frac{1}{2}} \: {\rm ctg} \: \pi \:  (1 - x)^{\frac{1}{2}}       
\end{equation}

\bigskip
\noindent
(we set here $\zeta = 0$ for simplicity)

\bigskip \bigskip
Finally, $\langle \tau^\mu_\mu \rangle$ will be given by  \\

\begin{eqnarray}
\langle \tau^\mu_\mu \rangle & = & - \left( \frac{\hbar H^3}{4 \pi c^2} \right) 
\frac{\int^{x_2}_{x_1} F(x) \rho (x) dx}{\int^{x_2}_{x_1} \rho (x) dx}   \nonumber \\    
& \equiv & - \left( \frac{\hbar H^3}{4 \pi c^2} \right) \frac{I_N}{I_D}
\end{eqnarray}

\bigskip
\noindent
where $x_1 = \bar{m}^2_0 \gamma$ . 
In our case, the adimensional variable $x$ runs in the interval $[ x_1 , \frac{3}{4})$ .
About the upper limit $x_2$ , a word on $F(x)$ is now in order.
$F(x)$ is a non-singular (monotonically) decreasing function in the interval $[0 , 1]$, and $F(x) > 0$
for $x$ in $[0 , \frac{3}{4} )$ . But this later interval is in fact the safe range for the physical
validity of $\langle T^\mu_\mu \rangle_{\rm QFT}$ (the mass of the test particle $m$ is much smaller 
than the mass scale $M_H$ of de Sitter Universe), and hence for ${\tilde {\langle T^\mu_\mu \rangle}}$ . 

\bigskip
On the other hand, if we consider the integral $I_N$ in the numerator of [Eq. (48)], the exponential of
$\rho (x)$ [Eq. (45)] plays a leading role in the interval $[x_1 , \frac{3}{4} )$ from the physical 
point of view since $\gamma^{-1} \gg 1$ . Therefore, the monotonically decreasing behavior of the function
$F(x)$ can be approximated by the straight line $y = - \frac{8}{3\pi} \left( x - \frac{3}{4} \right)$~.

\bigskip \bigskip
After a straightforward calculation one obtains for $I_N$ and $I_D$ [Eq. (48)] the following 
expressions  \\

\begin{equation}
I_N = - \frac{32}{3\pi} \left\lbrace - \frac{3}{4} \left\lbrack - \frac{e^{\frac{z}{\lambda}}}{2 z^2}
- \frac{e^{\frac{z}{\lambda}}}{2 \lambda z} + \frac{1}{2 \lambda^2} E_i \left( \frac{z}{\lambda} 
\right) \right\rbrack  - e^{\frac{z}{\lambda}} \left( z \lambda - \lambda^2 \right) + 2 E_i \left(
\frac{z}{\lambda} \right) \right\rbrace^{z_2}_{z_1}                       
\end{equation}

\bigskip
\noindent
and
\bigskip

\begin{equation}
I_D = 4 \left\lbrace - \frac{e^{\frac{z}{\lambda}}}{2 z^2} - \frac{e^{\frac{z}{\lambda}}}{2 \lambda z}
+ \frac{1}{2 \lambda^2} E_i \left( \frac{z}{\lambda} \right) \right\rbrace^{z_2}_{z_1}      
\end{equation}

\bigskip
\noindent
where

\begin{eqnarray}
z & = & \left\lbrack 1 - (1 - x)^{\frac{1}{2}} \right\rbrack^{\frac{1}{2}}    \\   
\lambda & = & \frac{\sqrt{3\gamma}}{2\pi}                                          
\end{eqnarray}

\bigskip
Considering the $\lambda ( \sqrt{\gamma} )$ leading terms, we have for $I_N$ and $I_D$~:

\begin{eqnarray}
I_N & \simeq & \frac{128}{3\pi} e^{\frac{2\pi}{\sqrt{6\gamma}}} \lambda^2 (1 + 7 \sqrt{2} \lambda )
\\  
I_D & \simeq & 16 e^{\frac{2\pi}{\sqrt{6\gamma}}} \lambda \left( \frac{1}{\sqrt{2}} + 3\lambda \right)
\end{eqnarray}       

\bigskip
\bigskip
From [Eq. (48)], [Eq. (53)] and [Eq. (54)], $\langle \tau^\mu_\mu \rangle $ reads, up to order $\gamma$
[Eq. (28)]~:  \\

$$\langle \tau^\mu_\mu \rangle = - \frac{\hbar H^3}{3\pi^3 c^2} \sqrt{6\gamma} \left( 1 + \frac{2}{\pi}
\sqrt{6 \gamma} \right)       \eqno(55.a)  $$

\bigskip
\bigskip
\noindent
Or, in terms of the scalar curvature $R \left\lbrack {\rm Eq.} (2b) ; R = 6 H^2 c^{-2} \right\rbrack$~:

$$\langle \tau^\mu_\mu \rangle = -\frac{R^2}{36\pi^3} \left( \frac{5\alpha^\prime c \hbar^3}{6}
\right)^{\frac{1}{2}} \left\lbrack 1 + \frac{2}{\pi} \left( \frac{5 \alpha^\prime \hbar}{4c} R 
\right)^{\frac{1}{2}} \right\rbrack       \eqno(55.b)  $$

\bigskip
Inserting $\langle \tau^\mu_\mu \rangle$ [Eq. (55.b)] into the back reaction [Eq. (24.a)] for $D = 3$ ,
we have

$$R - 6 \Lambda = \frac{4G R^2}{9 \pi^2 c^4} \left( \frac{5 \alpha^\prime \hbar^3 c}{6} 
\right)^{\frac{1}{2}} \left\lbrack 1 + \frac{2}{\pi} \left( \frac{5\alpha^\prime \hbar}{4c} R
\right)^{\frac{1}{2}} \right\rbrack      \eqno(56.a)  $$

\bigskip
\noindent
(For $D=3$ , $[G] = L^2 t^{-2} M^{-1}$)

\bigskip
\noindent
Or, [Eq. (24.b)] and [Eq. (24.c)],~:
\bigskip

$$ R = 6 \Lambda_{\rm eff}       \eqno(56.b)  $$

\noindent
where
\bigskip

\setcounter{equation}{56}
\begin{equation}
\Lambda_{\rm eff} = \Lambda + \frac{2 G R^2}{27 \pi^2 c^4} \: \left( \frac{5 \alpha^\prime \hbar^3 c}
{6} \right)^{\frac{1}{2}} \left\lbrack 1 + \frac{2}{\pi} \left( \frac{5 \alpha^\prime \hbar}{4c} R
\right)^{\frac{1}{2}} \right\rbrack                 
\end{equation}

\bigskip
\bigskip \bigskip
We are going to analyse now the physical consequences of the back reaction [Eq. (56.a)]. For simplicity
we consider $\langle \tau^\mu_\mu \rangle$ [Eq. (55.a)] up to order $\sqrt{\gamma}$ . We have

\bigskip

\begin{equation}
R - 6 \Lambda \simeq \hat{\alpha} R^2                 
\end{equation}

\noindent
where
\bigskip

\begin{equation}
\hat{\alpha} \equiv \frac{4G}{9 c^4 \pi^2} \left( \frac{5 \alpha^\prime c \hbar^3}{6} \right)^{\frac{1}
{2}}                        
\end{equation}

\bigskip
\noindent
[Eq. (58)] is a second order equation in $R$ , similar to the one found for the back reaction of 
massless quantum fields (including the graviton) in $4 -$ dimensional de Sitter space-time Ref. [7].
The exprssion for $\hat{\alpha}$ is here different as it contains $\alpha^\prime$ . (In the case of 
massless ${\rm QFT}$~, $\hat{\alpha}$ arise from the trace anomaly $\langle T^\mu_\mu \rangle$).

From [Eq. (58)] we have two solutions [Eq. (2.b)]

\bigskip
\begin{equation}
R_\pm = 6 \Lambda_\pm               
\end{equation}

\noindent
with
\bigskip

\begin{equation}
\Lambda_\pm = \frac{1}{12 \hat{\alpha}} \left\lbrack 1 \pm \left( 1 - 24 \Lambda \hat{\alpha} 
\right)^{\frac{1}{2}} \right\rbrack            
\end{equation}

\bigskip
\noindent
$\Lambda_\pm$ are the effective cosmological constants.

\bigskip \bigskip
One can distinguish three cases

\bigskip
\bigskip
(i)  If $\Lambda < \frac{1}{24 \hat{\alpha}} = \frac{3 c^4 \pi^2}{32 G \hbar} \left(
\frac{6}{5 \alpha^\prime c \: \hbar^3} \right)^{\frac{1}{2}}$

\bigskip
\noindent
We have two de Sitter space-times with curvatures [Eq. (60)]

\bigskip
\begin{equation}
R_\pm = \frac{9 c^4 \pi^2}{8 G} \: \left( \frac{6}{5 \alpha^\prime c \hbar^3} \right)^{\frac{1}{2}} \:
\left\lbrace 1 \pm \left\lbrack 1 - \frac{32 G \Lambda}{3 c^4 \pi^2} \: \left( \frac{5 \alpha^\prime
c \: \hbar^3}{6} \right)^{\frac{1}{2}} \right\rbrack^{\frac{1}{2}} \right\rbrace       
\end{equation}

\bigskip
\noindent
Both branches are well defined and have $R_\pm > 0$.

\bigskip \bigskip
(ii)  If  $\Lambda = \frac{1}{24 \hat{\alpha}} $~, there is a unique de Sitter space-time

\bigskip
\begin{equation}
R_+ = R_- = \frac{1}{4 \hat{\alpha}} = \frac{9 c^4 \pi^2}{16 G} \: \left( \frac{6}{5 \alpha^\prime 
\hbar^3 c} \right)^{\frac{1}{2}}                              
\end{equation}

\bigskip
\bigskip
(iii) If  $\Lambda > \frac{1}{24 \hat{\alpha}}$~, there are neither physical real curvatures nor 
temperatures.

\bigskip
For small $\Lambda$~,  

$$\Lambda \ll \frac{1}{24} \hat{\alpha} \left( = \frac{3 c^4 \pi^2}{32 G} \: \left( \frac{6}{5 c
\alpha^\prime \hbar^3} \right)^{\frac{1}{2}} \right) \: , $$

\noindent 
for which $\langle T^\mu_\mu \rangle_{\rm QFT}$ - and hence ${\tilde {\langle T^\mu_\mu \rangle}}$
and $\langle \tau^\mu_\mu \rangle$ - are not trivial, we have

\bigskip
$$ R_- \simeq 6 \Lambda \ll \frac{1}{4 \hat{\alpha}}        \eqno(66.a)  $$

\bigskip
$$R_+ \simeq \frac{1}{\hat{\alpha}} = \frac{9 c^4 \pi^2}{4 G} \: \left( \frac{6}{5 \alpha^\prime c \:
\hbar^3} \right)^{\frac{1}{2}}   \equiv R_{\max}       \eqno(66.b)   $$

\bigskip \bigskip
From the above equations we see that one recovers the classical space-time for the $R_-$
solution. $R_-$ is a small curvature solution. On the contrary, $R_+$ does not represent a classical 
allowed configuration and its curvature is very high. 
The two branches of solutions are generically of different kind. We call the $R_-$ branch  
``classical'' as it represents solutions which are classically allowed, while the $R_+$ branch will be
 the ``quantum'' one as the configurations do not occur classically. \\

\bigskip
From [Eqs. (60)] and [Eq. (61)], we read a maximum value for the effective cosmological constant~:

\bigskip
\setcounter{equation}{66}
\begin{equation}
\Lambda_{\max} \simeq \frac{1}{6 \hat{\alpha}} = \frac{3 c^4 \pi^2}{8 G} \: \left( \frac{6}{5 
\alpha^\prime c \: \hbar^3} \right)^{\frac{1}{2}}              
\end{equation}

\bigskip
In terms of $\Lambda_{\max}$ (alternatively of $R_{\max}$~, we have

\bigskip
\begin{equation}
\Lambda_{\pm} = \frac{1}{2} \Lambda_{\max} \lbrack 1 \pm \left( 1 - 4 \: \frac{\Lambda}{\Lambda_{\max}}
\right)^{\frac{1}{2}} \rbrack       
\end{equation}

\bigskip
\begin{equation}
R_{\pm} = \frac{1}{2} R_{\max} \lbrack 1 \pm \left( 1 - 4 \: \frac{R}{R_{\max}} \right)^{\frac{1}{2}}
\rbrack       
\end{equation}

\bigskip
\bigskip
In an expansion in $\frac{R}{R_{\max}}$, the leading order is $R_{(-)} = R$, $R_{(+)} = R_{\max}$. QFT 
de Sitter temperature [Eq. (6.b)] associated to the classical branch $R_{(-)}$ is

\bigskip
$$T_{-DS} = \frac{\hbar c}{2 \pi k_B} \: \left( \frac{R_{(-)}}{6} \right)^{\frac{1}{2}}$$

\bigskip
The string quantum branch $R_{(+)}$ has a string temperature

\bigskip
$$T_{+ {\rm string}} = \frac{c^2}{\alpha^{\prime} k_B} \: \left( \frac{6}{R_{(+)}} 
\right)^{\frac{1}{2}} $$

\bigskip \bigskip \bigskip

\section{CONCLUSIONS}

\bigskip \bigskip \bigskip
Combined study of QFT and string theory in curved backgrounds allowed us to go further in the
understanding of quantum gravity effects. \\

\noindent
The string analogue model (or thermodynamical approach) is a suitable framework in cosmology and black
holes to combine both QFT and string study, and address the problem of quantum string back reaction. \\

\noindent
The Dual relationship shown here between the two domains: classical-QFT and quantum string, applies 
also to other space-times and plays a key role in the black hole case Ref. [5]. \\

\noindent
String black hole temperature and quantum string back reaction for black holes is reported in another
paper Ref. [5]. The two phases correspond to the evaporation from a classical black hole geometry with
intrinsic temperature given by the QFT Hawking temperature to a string phase for the geometry 
(sustained by the quantum string back reaction) which temperature becomes the intrinsic string
temperature. \\

\noindent
These studies and our Dual relation between classical-QFT and string phases appear irrespective of 
conformal invariance. \\

\noindent
Similar study for Anti de Sitter space time is under investigation by these authors. \\

\noindent
QFT in Anti de Sitter space time does not possess an intrinsic or Hawking temperature. Strings in
AdS space-time do not possess a maximal or critical temperature neither Ref. [3] and [4]. The 
partition function for a gas of strings in AdS space-time is defined at any positive temperature
Ref. [3]. \\

\noindent
Such results for strings in AdS space-time were also confirmed  in the presence of a full conformal
invariant AdS string background WZWN model SL$(2, R)$ (AdS with torsion) Ref. [16]. As shown in 
Ref. [16] conformal invariance {\it simplifies} the mathematics of the problem but the physics remain
mainly {\it unchanged}. For low and high masses, the string mass spectra in conformal and non conformal
backgrounds are the same. \\

\noindent
The purpose of this paper was to go further in the understanding of string theory in de Sitter 
space-time. Motivate (and {\it at priory} justify) the choice of de Sitter space time: $(i)$ the
cosmological (inflationary) relevance of de Sitter space-time, $(ii)$ the present knowlodge of string 
dynamics in conformal and non conformal invariant backgrounds, in particular in the conformal and non
conformal invariant AdS backgrounds above mentioned, $(iii)$ the lack, at the present time, of a full
string conformal invariant treatement involving de Sitter space-time.

\bigskip

{\Large \underline{Acknowledgements}}

\bigskip \bigskip \bigskip
M.R.M. acknowledges the Spanish Ministry of Education and Culture (D.G.E.S.) for partial financial 
support (under project~: ``Plan de movilidad personal docente e investigador'') and the Observatoire 
de Paris - DEMIRM for the kind hospitality during this work.

\bigskip
Partial financial support from NATO Collaborative Grant CRG $974178$ is also acknowledged.

\newpage

{\Large \underline{References}}

1. G.W. Gibbons and S.W. Hawking, Phys. Rev. D15 (1977) 2738.
\medskip

2. H.J. de Vega, N. S\'anchez, Phys. Lett. B197 (1987) 320.
\medskip

3. A.L. Larsen, N. S\'anchez, Phys. Rev. D52 (1995) 1051.
\medskip

4. H.J. de Vega, A.L. Larsen, N. S\'anchez, Phys. Rev. D51 (1995) 6917.
\medskip

5. M. Ramon Medrano and N. S\'anchez ``{\it Hawking Temperature, String Temperature and the String 
Phase of Black Hole Spacetime}'', manuscript in preparation.
\medskip

6. S. Wada and T. Azuma, Phys. Lett. 132B (1983) 313.
\medskip

7. M.A. Castagnino, J.P. Paz and N. S\'anchez, Phys. Lett. B193 (1987).
\medskip

8. T.S. Bunch and P.C.W. Davies, Proc. R. Soc. London A360 (1978) 117.
\medskip

9. See for example G.W. Gibbons in ``{\it General Relativity, An Einstein Centenary Survey}'', 
Eds. S.W. Hawking and W. Israel, Cambridge University Press, UK (1979). \\

N.D. Birrell, P.C.W. Davies ``{\it Quantum Fields in Curved Space}'', Cambridge Universsity Press,
UK (1982).
\medskip

10. H.J. de Vega and N. S\'anchez, Phys. Rev. D47 (1993) 3394.
\medskip

11. See for example B.S. De Witt, in ``{\it General Relativity, An Einstein Centenary Survey}'', Eds. 
S.W. Hawking and W. Israel, Cambridge University Press, UK (1979).
\medskip

12. H.J. de Vega, A.L. Larsen, N. S\'anchez, Nucl. Phys. B427 (1994) 643.
\medskip

13. J.S. Dowker, R. Critchley, Phys. Rev. D13 (1976) 3224. 
\medskip

14.S.M. Christensen, Phys. Rev. D17 (1978) 946. 
\medskip

15. S.M. Christensen, M.J. Duff, Nucl. Phys. B170 (1980) 480.
\medskip

16. H.J. de Vega, A.L. Larsen and N. S\'anchez, Phys. Rev. 58D (1998) 026001.

\end{document}